\documentclass[natbib]{svjour3}

\usepackage[english]{babel}
\usepackage{amssymb,amsmath}
\usepackage{mathptmx}

\newcommand{\pref}[1]{(\ref{#1})}
\newcommand{\Sec}[1]{Sect.~\ref{#1}}
\newcommand{\Eq}[1]{Eq.~(\ref{#1})}

\DeclareMathOperator*{\eins}{\mathbf{1}}
\DeclareMathOperator*{\tr}{tr}
\DeclareMathOperator*{\D}{d}

\journalname{International Journal of Theoretical Physics}

\bibpunct{(}{)}{,}{}{}{;}

\begin{document}


\title{Epistemic Entanglement due to Non-Generating Partitions of Classical Dynamical Systems}
\titlerunning{Epistemic entanglement}

\author{Peter beim Graben
    \and
    Thomas Filk
    \and
    Harald Atmanspacher
 }

\authorrunning{beim Graben et al.}

\institute{
    Peter beim Graben \at
    Dept. of German Language and Linguistics, and
    Bernstein Center for Computational Neuroscience,
    Humboldt-Universit\"at zu Berlin,
    Unter den Linden 6,
    D-10099 Berlin,
    \email{peter.beim.graben@hu-berlin.de}
    \and
    Thomas Filk \at
    Physikalisches Institut,
    University of Freiburg, Germany
    \and
    Harald Atmanspacher \at
    Institut f\"ur Grenzgebiete der Psychologie und Psychohygiene,
    Freiburg i. Br., Germany
 }

\date{\today}
\maketitle

\begin{abstract}
Quantum entanglement relies on the fact that pure quantum states are dispersive and often inseparable. Since pure classical states are dispersion-free they are always separable and cannot be entangled. However, entanglement is possible for epistemic, dispersive classical states. We show how such epistemic entanglement arises for epistemic states of classical dynamical systems based on phase space partitions that are not generating. We compute epistemically entangled states for two coupled harmonic oscillators.
\end{abstract}


\section{Introduction}
\label{sec:intro}

Entanglement is a well-known and central concept in quantum theory \citep{Schrodinger35a, EinsteinPodolskyRosen35}, where it expresses a fundamental nonlocality (holism) of ontic quantum states, regarded as independent of epistemic means of gathering knowledge about them. In quantum theory, entanglement is tightly related to other key features such as dispersive ontic states, non-commuting observables, and an orthomodular lattice structure, giving rise to incompatible and even complementary descriptions of the behavior of quantum systems.

Some of the pioneers of quantum theory, notably Bohr and Pauli, were convinced that the ``situation in regard to complementarity within physics leads naturally beyond the narrow field of physics to analogous situations in connection with the general conditions of human knowledge'' \citep{Pauli50}. This conjecture has indeed been picked up by a number of research groups in quantum logics \citep{Foulis99}, psychology and cognitive science (cf.~\citet{Atmanspacher11}, Sec.~4.7 for compact reviews), where they resolved old puzzles and yielded new insights for specific situations.

Given these successful applications, the question arises whether the applicability of concepts, which were viably introduced in quantum theory, to non-quantum systems can be understood in a more general way. In this respect, a number of researchers investigated in detail how the concept of complementarity can be sensibly addressed in classical dynamical systems as represented in a suitable phase space. The formal key to such a generalized version of complementarity lies in the construction of coarse-grainings by state space partitions giving rise to either orthomodular lattice structures of the resulting propositional logics \citep{DvurecenskijPulmannovaSvozil95, Foulis99, Svozil05, WrightR90}, or contextual probabilities \citep{Khrennikov03a, Khrennikov10}. More recently, \citet{GrabenAtmanspacher06a, GrabenAtmanspacher09} constructed dispersive epistemic states from phase space partitions. Descriptions based on partitions are compatible only under very specific conditions, otherwise they are incompatible or complementary \citep{DvurecenskijPulmannovaSvozil95, GrabenAtmanspacher06a, GrabenAtmanspacher09, Khrennikov03a, Svozil05}. More precisely, it turns out that compatible descriptions arise only for generating partitions, a concept introduced into ergodic theory in seminal work by \citet{Kolmogorov58} and \citet{Sinai59}.

The pivotal hallmark of generating partitions is that their construction implements the dynamics of the system considered, so that the epistemic states represented by the partition cells evolve in a way which is topologically equivalent to the underlying dynamics. This important fact is crucial for the entire field of symbolic dynamics \citep{LindMarcus95}, which studies the behavior of complex systems in terms of symbolic descriptions based on generating partitions.

Non-generating (also called improper) partitions are usually considered detrimental in symbolic dynamics \citep{BolltStanfordEA01}. However, they allow us to understand better how incompatibility and complementarity can arise in classical dynamical systems (which are increasingly used to model systems in cognitive science, cf.~\citet{Gelder98}). \Citet{GrabenAtmanspacher09} even suggested how to extend this idea to the notions of comparability and commensurability.

In quantum theory complementarity is closely related to the concept of entanglement, both concepts relying on the dispersion of ontic quantum states. For stochastic dynamical systems, the emergence of entanglement under coarse-grainings has been proven by \citet{AllahverdyanKhrennikovNieuwenhuizen05}. In a recent paper \citep{AtmFilkGrab11} we conjectured that complementarity due to dispersive epistemic states based on non-generating partitions may entail a similarly defined concept of entanglement. In the present study, we demonstrate in detail that and how such an epistemic entanglement is indeed feasible.

The following Section 2 recalls how entanglement is fundamentally based on the inseparability of pure states of quantum systems (2.1). In classical systems, pure states are always separable and thus not entangled (2.2). Section 3 addresses the option of entanglement for epistemic classical states based on phase space partitions (3.1) and illustrates this for a harmonic oscillator. Section 4 discusses the resulting epistemic entanglement for two coupled harmonic oscillators. Section 5 offers some conclusions and perspectives.

\section{Separability and Entanglement}
\label{sec:sepent}

In this section we shall briefly recall the notions of entanglement and separability in quantum  physics and classical physics. In both cases, \emph{states} can be defined as linear, positive, and normalized functionals on an algebra of observables. \emph{Pure states} are distinguished states on the boundary of the state space.  Pure states cannot be decomposed into other states;  they are dispersion-free with respect to a maximal algebra of commuting observables. In classical physics  this includes all observables but in quantum physics such a maximal algebra of commuting observables is only a subset of all observables.

\subsection{Separability and Entanglement in Quantum Systems}
\label{sec:seqt}

In quantum mechanics, the state space of a composed system is the tensor product of the state spaces of its components  A, B. Therefore, in a Hilbert space representation, we have $\mathcal{H} = \mathcal{H}_{\rm A} \otimes \mathcal{H}_{\rm B}$. Let $\{|a_i \rangle\ | i \in \mathbb{N}\}$ and $\{|b_j \rangle\ | j \in \mathbb{N} \}$ be a complete orthonormal basis of $\mathcal{H}_{\rm A}$ and $\mathcal{H}_{\rm B}$, respectively. Then, a generic \emph{pure state} in $\mathcal{H}$ can be written as
\begin{equation}\label{eq:purest}
   |\psi \rangle = \sum_{ij} c_{ij} |a_i \rangle\ \otimes |b_j \rangle\ \:.
\end{equation}

The state $|\psi \rangle$ is called \emph{separable}, if the coefficients $c_{ij}$ factorize: $c_{ij} = c^{\rm A}_i c^{\rm B}_j$. In this case, also the state $|\psi \rangle$ factorizes into a tensor product of pure component states
\begin{equation}\label{eq:puresep}
 |\psi \rangle = |a \rangle \otimes |b \rangle \:,
\end{equation}
with $|a \rangle = \sum_i c^{\rm A}_i |a_i \rangle$ and $|b \rangle = \sum_j c^{\rm B}_j |b_j \rangle$. A pure state which is not separable is called an \emph{entangled state}. This definition does not depend on the chosen basis.

In quantum  physics, a state can generally be represented by a Hermitian trace-class operator, the density matrix $\rho$, acting on $\mathcal{H}$. Using density matrices, the expectation value functional is given by $\langle Q \rangle_\rho = \tr (\rho Q)$. The density matrix for a pure state $|\psi \rangle \in \mathcal{H}$ is the projector $\rho_\psi = |\psi \rangle \langle \psi|$.

Let
\begin{equation}\label{eq:puredm}
 \rho_\psi = (|a \rangle \otimes |b \rangle) (\langle a | \otimes \langle b |)
 =   |a \rangle \langle a | \otimes |b \rangle \langle b |
\end{equation}
be the density matrix corresponding to the separable pure state $|\psi \rangle$ of \Eq{eq:puresep}. Performing the trace over the local Hilbert space $\mathcal{H}_{\rm B}$ yields
\begin{eqnarray*}
  \tr{}_{\rm B} (\rho_\psi) &=& \sum_m (\eins \otimes \langle b_m |)
  \rho_\psi
  (\eins \otimes |b_m \rangle) \\
  &=& \sum_m (\eins \otimes \langle b_m |)
  (|a \rangle \langle a | \otimes |b \rangle \langle b |)
  (\eins \otimes |b_m \rangle) \\
  &=& \sum_m |a \rangle \langle a | \langle b_m |b \rangle \langle b | b_m \rangle \\
  &=& |a \rangle \langle a | \sum_m |b_m|^2 \\
  &=& |a \rangle \langle a | \:,
\end{eqnarray*}
which is a pure state again. However, for an entangled state $|\psi \rangle$, the density matrix resulting from the partial trace corresponds to a  statistical state (or mixed state, briefly  mixture) in the local Hilbert space $\mathcal{H}_{\rm A}$. The  intuition behind this statement is that  the partial trace of an entangled state leads to a loss of information about this state due to quantum correlations between the two systems  --- which is the hallmark of a mixed state.

In \Sec{sec:epent}, we shall  generalize this observation to classical systems, in particular to classical dynamical systems.

\subsection{Separability and Entanglement in Classical Systems}
\label{sec:sesm}

In classical  physics, statistical states are  represented by probability measures over the phase space $X \subset \mathbb{R}^d$, where $d \in \mathbb{N}$ denotes the dimensionality of the system. Thus, we need a probability space $\Omega = (X, \mathcal{B}, \mu)$ with a Borel $\sigma$-algebra $\mathcal{B}$ of measureable sets from $X$ and a reference measure $\mu : \mathcal{B} \to [0, 1]$ \citep{Walters81}.
Suppose that the measure $\mu$ is absolutely continuous,  so that any statistical state $\rho$ can be represented by its associated probability density function (p.d.f.) $p_\rho(x)$ over phase space $X$ according to
\begin{equation}\label{eq:classexpec}
 \langle f \rangle_\rho = \int_X f(x) p_\rho(x) \D \mu(x) \:,
\end{equation}
where $f: X \to \mathbb{R}$ is a classical observable over phase space $X$.

If the system can be decomposed into subsystems A, B, the phase space can be written as a Cartesian product $X = X_{\rm A} \times X_{\rm B}$. Let $p_{\rm A}$ and $p_{\rm B}$ be p.d.f.s over $X_{\rm A}$ and $X_{\rm B}$, respectively. Then the state $\rho$ is \emph{separable} if and only if
\begin{equation}\label{eq:statsepar}
  p_\rho(x) = p_{\rm A}(x_{\rm A}) p_{\rm B}(y_{\rm B})
\end{equation}
where $x = (x_{\rm A}, x_{\rm B})$ is the ordered pair of points $x_{\rm A} \in X_{\rm A}$ and $x_{\rm B} \in X_{\rm B}$.

In this picture, pure states are  represented by Dirac measures $\delta_x: \mathcal{B} \to \{0, 1\}$  over points $x \in X$ in phase space, such that for any $A\in \mathcal{B}$ we have $\delta_x(A) = 1$ if $x \in A$ and $\delta_x(A)=0$ otherwise \citep{Walters81}.  Dirac measures are always separable by means of the factorizability of Dirac's delta functions
\[
 \delta(x) =   \delta(x_{\rm A}) \delta(x_{\rm B}) \:.
\]
This implies that pure states in classical  systems are always separable. This is also reflected by the fact that observables $f$ are evaluated pointwise in phase space, i.e. $f(x)$ is the result of evaluating $f$ in the pure state $x \in X$. Thus, in classical  systems pure states can never be entangled with respect to the full observable algebra.

\section{Epistemic Quantization of Classical Dynamical Systems}
\label{sec:eqds}

Pure states of classical systems, represented as points $x \in X$ or, alternatively, as Dirac measures over $X$, are also called \emph{ontically pure} \citep{AtmanspacherPrimas03}. This notion expresses that they refer to the limiting concept of ``idealized'' states, regarded as independent of their observation or knowledge about them. While the no-go statement for entangled states in classical systems refers to such ontically pure states, we will now demonstrate that it may be invalid from an epistemic point of view.

\Citet{GrabenAtmanspacher06a, GrabenAtmanspacher09} have recently shown that classical dynamical systems may exhibit incompatible (non-commuting) properties when the state space is epistemically restricted by observational coarse-grainings, in close analogy to the ``firefly box'' and ``generalized urn models'' of \citet{Foulis99} and \citet{WrightR90}, respectively. In this case it is possible that statistical states cannot be refined by the observable algebra and have some residual finite phase space grain as support. For such states one can find pairs of classical observables that cannot be simultaneously measured with precise accuracy. Thus, such observables  are indeed incompatible.

Residual grains in epistemic coarse-grainings can lead to entanglement if particular statistical states are distinguished as \emph{epistemically pure} with respect to a certain coarse-graining. If a state is epistemically pure with respect to a global observable (or a set of global observables), this state may appear as a mixture with respect to a local observable. In other words, the partical trace of such an epistemically pure state may lead to a  mixture. Corresponding epistemically pure states will be called \emph{epistemically entangled}.

\subsection{Phase Space Partitions}
\label{sec:psp}

Consider a classical discrete-time and invertible dynamical system $(X, \Phi)$ with phase space $X$ and flow $\Phi$, generating the dynamics by successive iterations $x_t = \Phi^t(x_0)$,  with initial condition $x_0 \in X$ and parameter time $t \in \mathbb{Z}$. As mentioned above, physical observables are real-valued functions $f: X \to \mathbb{R}$ assuming a value $a = f(x)$ in state $x \in X$.  Statistical states are  represented by p.f.d.s $p_\rho: X \to \mathbb{R}_0^+$ such that \Eq{eq:classexpec} gives the expectation value of $f$ in state $\rho$. Inserting $f(x) = 1$ into \pref{eq:classexpec} gives the normalization condition for $p_\rho$.

Individual observables $f$ are generally not injective. Therefore, many states within a set $A \subset X$ are simultaneously evaluated with the same measurement result $a = f(x)$ by $f$. We call two states $x, y \in X$ \emph{epistemically equivalent with respect to $f$} if
\begin{equation}\label{eq:epequiv}
    f(x) = f(y)
\end{equation}
and denote this symbolically by $x \overset{f}{\sim} y$ \citep{GrabenAtmanspacher06a, GrabenAtmanspacher09}. This equivalence relation entails a partition (which we assume to be finite) $\mathcal{P}_f = \{ A_1, A_2, \dots A_n\}$ of $X$ into $n \in \mathbb{N}$ pairwise disjoint sets such that
$A_i \cap A_j = \emptyset$ for all $i \ne j$ and $\bigcup_i^n A_i = X$.

Given two partitions $\mathcal{P} = \{ A_1, A_2, \dots A_n\}$ and $\mathcal{Q} = \{ B_1, B_2, \dots B_m\}$, the product partition $\mathcal{P} \vee \mathcal{Q}$ is defined by the partition containing all intersections of cells from $\mathcal{P}$ with cells of $\mathcal{Q}$, i.e.
\begin{equation}\label{eq:prodpart}
 \mathcal{P} \vee \mathcal{Q} = \{ A_i \cap B_j |
        A_i  \in \mathcal{P}, B_j  \in \mathcal{Q}\}
\end{equation}
In general, the product partition is a refinement of both partitions $\mathcal{P}$ and $\mathcal{Q}$.

In order to describe successive measurements of an observable $f$ over a period of time $T$, one has to combine the  evaluation of $f$ with the dynamics $\Phi$. Starting with the initial condition $x_0 \in X$ at time $t = 0$, a  measurement of $f$ yields the information $x_0 \in A_{i_0}$ for one distinguished cell $A_{i_0} \in \mathcal{P}_f$. After a single time step a repeated measurement of $f$ in the subsequent state $x_1 = \Phi(x_0) \in X$ yields $x_1 \in A_{i_1}$ ($A_{i_1} \in \mathcal{P}_f$), or, equivalently, $x_0 \in \Phi^{-1}(A_{i_1})$. Combining both measurement results gives $x_0 \in A_{i_0} \cap \Phi^{-1}(A_{i_1})$, which is a cell of the product partition $\mathcal{P}_f \vee \Phi^{-1}(\mathcal{P}_f)$, where
\begin{equation}\label{eq:preimage}
    \Phi^{-1}(\mathcal{P}) = \{ \Phi^{-1}(A_i) | A_i \in \mathcal{P} \}
\end{equation}
is the pre-image of partition $\mathcal{P}$ under the flow $\Phi$. The \emph{dynamic refinement} of a partition $\mathcal{P}$ over a period of time $T$ is then defined by the product partition
\begin{equation}\label{eq:refine}
 \bigvee_{t = 0}^T \Phi^{-t}(\mathcal{P}) \:.
\end{equation}

In the limit of continuous measurements beginning in the infinite past and terminating in the infinite future, the \emph{finest dynamic refinement} is given by
\begin{equation}\label{eq:finestrefine}
    \mathbf{R} \mathcal{P} =
   \bigvee_{t = -\infty}^\infty \Phi^t(\mathcal{P})\, ,
\end{equation}
where $\mathbf{R}$ is the ``finest-refinement operator'' acting on a partition $\mathcal{P}$ \citep{GrabenAtmanspacher09}.  Ideally, such continuous measurements yield complete information about the initial condition $x_0$ in phase space. This is achieved if the finest dynamical refinement \Eq{eq:finestrefine} entails the so-called identity partition,
\begin{equation}\label{eq:generator}
    \mathbf{R} \mathcal{P} = \mathcal{I} \:,
\end{equation}
where $\mathcal{I} = \{I_x = \{  x \} |  x \in X \}$ consists  exclusively of singletons of phase space points. A partition $\mathcal{P}$ obeying \Eq{eq:generator} is called \emph{generating}  \citep{Corn82}. If $\mathcal{P}$ is not generating, its cells $P\in \mathcal{P}$ are non-singleton sets with some  residual grain.

From the finest dynamic refinement
\begin{equation}\label{eq:fref}
    \mathcal{F} = \mathbf{R} \mathcal{P}_f
\end{equation}
of a partition $\mathcal{P}_f$ induced by an epistemic observable $f$, one resumes the description of continuous measurements of arbitrary finite duration $T$ [\Eq{eq:refine}] by joining subsets of $\mathcal{F}$ that are visited by the system's trajectory during measurement.  This is  formally expressed by the concept of a \emph{partition algebra} $\mathcal{B}_f = \mathcal{A}(\mathcal{F})$ of $\mathcal{F}$, which is the smallest Borel $\sigma$-subalgebra of $\mathcal{B}$ (given the probability space $\Omega$ from \Sec{sec:sesm}) containing $\mathcal{F}$. Then, we call a statistical state $\rho$ \emph{epistemically accessible with respect to the observable $f$} (or $f$-e.a.) if its p.d.f. $p_\rho$ is measurable in $\mathcal{B}_f$ \citep{GrabenAtmanspacher06a}.

Note that ``accessible with respect to the observable $f$'' may also mean ``accessible with respect to a set of observables $\{f\}$'', and this set can contain all observables which are actually realizable.  For the construction defined above it is relevant that this set of observables generates a  phase space partition $\mathcal{P}$. In this sense the finest dynamic refinement  with respect to all epistemically realizable observables will be called the ``set of {\it epistemically pure states}''. This  notion is motivated by the fact that no epistemically realizable observable is able to decompose these states into finer states.

If the epistemically realizable observables are only a subset of all mathematically possible observables (functions on phase space), the finest dynamical refinement will differ from $\mathcal{I}$.  Under certain conditions the resulting non-singleton cells can be regarded as supports of epistemically pure states and, with respect to a modified separability criterion, they can even be entangled.

\subsection{Example: Harmonic Oscillator}
\label{sec:ho}

Consider a harmonic oscillator with Lagrangian $L = \frac{1}{2}(\dot{q}^2 - \omega^2 q^2)$, where $\omega$ denotes the angular frequency of the  oscillator, $q$ the (generalized) position, and $p = \partial L / \partial \dot{q} = \dot{q}$ the canonically  conjugate momentum of $q$. The resulting Euler-Lagrange equation
\begin{equation}\label{eq:eulerlagrange}
    \ddot{q} + \omega^2 q = 0
\end{equation}
has the solutions
\begin{eqnarray}
 \label{eq:dglsolv1}
  q(t) &=& A \cos \omega t + B \sin \omega t \\
 \label{eq:dglsolv2}
  p(t) &=& - A \omega \sin \omega t + B \omega \cos \omega t \:.
\end{eqnarray}

Using initial conditions $q(0) =A$ and $p(0) = B \omega$ and introducing phase space coordinates $x = (q, p)^T \in X$, we obtain
\begin{equation}\label{eq:hosolve}
  x(t) = \Phi^t \cdot x(0)
\end{equation}
with the flow given by
\begin{equation}\label{eq:hoflow}
   \Phi^t =
   \begin{pmatrix}
     \cos \omega t & \frac{1}{\omega} \sin \omega t \\
     - \omega \sin \omega t & \cos \omega t \\
   \end{pmatrix} \:.
\end{equation}

In order to introduce a coarse-graining, we first have to discretize time. With $T= 2 \pi / \omega$ as the oscillator's period, let $\tau = T / 2$ be our sampling time. The sampled flow is then
\begin{equation}\label{eq:hoflo2}
   \Phi^\tau =
   \begin{pmatrix}
     -1 & 0 \\
     0 & -1 \\
   \end{pmatrix} = - \eins
\end{equation}
with unit matrix $\eins$. From \pref{eq:hoflo2}  it follows that $(\Phi^\tau)^2 = \eins$ and therefore $\Phi^{-\tau} = (\Phi^\tau)^{-1} = \Phi^\tau$.

Now we introduce two coarse-grained observables, one for position and one for momentum:
\begin{eqnarray}
 \label{eq:coarse-q}
    Q(x) &=& \Theta(q) \\
  \label{eq:coarse-p}
    P(x) &=& \Theta(p) \:,
\end{eqnarray}
with the Heaviside function
\begin{equation}\label{eq_heavi}
 \Theta(u) = \begin{cases}
 0 & : \quad u < 0 \\
 1 & : \quad u \ge 0 \:.
 \end{cases}
\end{equation}

These observables induce two phase space partitions: $\mathcal{Q} = \{A_0, A_1\}$ with $A_0 = \{  x \in X | Q( x) = 0 \}$, $A_1 = \{  x \in X | Q( x) = 1\}$, and $\mathcal{P} = \{B_0, B_1 \}$ with $B_0 = \{  x \in X | P( x) = 0 \}$, $B_1 = \{  x \in X | P( x) = 1 \}$.

Next, we determine the finest refinements \Eq{eq:finestrefine} of those partitions. Because $\Phi^{-\tau} \cdot x = - x$, we obtain $\Phi^{-\tau}(A_0) = A_1$, and vice versa, $\Phi^{-\tau}(A_1) = A_0$. Likewise, $\Phi^{-\tau}(B_0) = B_1$, and $\Phi^{-\tau}(B_1) = B_0$, and therefore
\begin{eqnarray}
    \label{eq:frfQ}
    \mathbf{R} \mathcal{Q} &=&
    \Phi^{-\tau}(\mathcal{Q}) \vee \mathcal{Q} = \mathcal{Q}
    \\
  \label{eq:frfp}
    \mathbf{R} \mathcal{P} &=&
    \Phi^{-\tau}(\mathcal{P}) \vee \mathcal{P} = \mathcal{P} \:.
\end{eqnarray}
Eventually, the associated partition algebras, i.e.~the smallest $\sigma$-algebras from $\mathcal{B}$ containing either $\mathcal{A}(\mathbf{R} \mathcal{Q})$ or $\mathcal{A}(\mathbf{R} \mathcal{P})$ are the algebras $\mathcal{B}_Q =\mathcal{A}(\mathcal{Q})$ and $\mathcal{B}_P = \mathcal{A}(\mathcal{P})$,
respectively.

Now, let $p_\rho(x) = 2 \chi_{A_1}(x)$ be a uniform p.d.f. over phase space $X$ supported by the partition cell $A_1 \in \mathcal{Q}$, where  $\chi_{A_1}$ is the characteristic function of $A_1$ with measure $\mu(A_1) = 0.5$ (because $\mathcal{Q}$ bi-partitions $X$ equally and because $\mu$ is normalized over $X$: $\mu(X) = 1$). The statistical state $\rho$ associated with $p_\rho$ is epistemically accessible with respect to the coarse-grained observable $Q$, since $p_\rho$ is $\mathcal{B}_Q$-measurable.

Moreover, the dispersion $\Delta_{\rho} (Q)$ of state $\rho$ with respect to the observable $Q$ can be computed from $\langle Q \rangle^2_\rho$ and $\langle Q^2 \rangle_\rho$:
\begin{eqnarray*}
\langle Q \rangle_\rho  &=&
    \int_X 2 \chi_{A_1}(x) Q(x) \D \mu(x) \\
&=& 2 \int_{A_1} Q(x) \D \mu(x) \\
&=& 2 \int_{A_1} 1 \D \mu(x) = 1
\end{eqnarray*}
\begin{eqnarray*}
\langle Q^2 \rangle_\rho  &=&
    \int_X 2 \chi_{A_1}(x) Q^2(x) \D \mu(x) \\
&=& 2 \int_{A_1} Q^2(x) \D \mu(x) \\
&=& 2 \int_{A_1} 1 \D \mu(x) = 1
\end{eqnarray*}
Since
\[   \Delta_\rho(Q) =
       \sqrt{\langle Q^2 \rangle_\rho -
\langle Q \rangle^2_\rho} = 0 \:, \]
$\rho$ is dispersion-free with respect to $Q$, and $\rho$ is an eigenstate of $Q$.

Likewise, the dispersion $\Delta_{\rho} (P)$ of state $\rho$ with respect to the coarse-grained observable $P$ can be computed from $\langle P \rangle^2_\rho$ and $\langle P^2 \rangle_\rho$:
\begin{eqnarray*}
\langle P \rangle_\rho  &=&
    \int_X 2 \chi_{A_1}(x) P(x) \D \mu(x) \\
&=& 2 \int_{A_1} P(x) \D \mu(x) \\
&=& 2 \left[ \int_{A_1 \cap B_0} P(x) \D \mu(x) +
 \int_{A_1 \cap B_1} P(x) \D \mu(x) \right] \\
&=& 2 \int_{A_1 \cap B_1} 1 \D \mu(x) \\
&=& 2 \mu(A_1 \cap B_1) = \frac{1}{2}
\end{eqnarray*}
\begin{eqnarray*}
\langle P^2 \rangle_\rho  &=&
    \int_X 2 \chi_{A_1}(x) P(x) \D \mu(x) \\
&=& 2 \int_{A_1} P^2(x) \D \mu(x) \\
&=& 2 \left[ \int_{A_1 \cap B_0} P^2(x) \D \mu(x) +
 \int_{A_1 \cap B_1} P^2(x) \D \mu(x) \right] \\
&=& 2 \int_{A_1 \cap B_1} 1 \D \mu(x) \\
&=& 2 \mu(A_1 \cap B_1) = \frac{1}{2}
\end{eqnarray*}
Since
\[
\Delta_\rho(P) = \sqrt{\langle P^2 \rangle_\rho - \langle P \rangle^2_\rho} =
\sqrt{\frac{1}{2} - \frac{1}{4}} = \frac{1}{2} \:,
\]
$\rho$ is not dispersion-free with respect to $P$ and hence not an eigenstate of $P$.  This shows that coarse-grained observables $Q$ and $P$ are incompatible for the sampled classical harmonic oscillator:  they do not have every eigenstate in common.

Note that a different sampling, e.g. $\tau =  T / 4$ rather than $\tau = T / 2$ yields a  different picture. In this case the finest refinements of $\mathcal{Q}$ and $\mathcal{P}$ are identical, $\mathcal{Q} \vee \mathcal{P}$, such that every state that is epistemically accessible by $Q$ is also epistemically accessible by $P$ (and vice versa),  so that $Q$ and $P$ turn out to be compatible.

\section{Epistemic Entanglement}
\label{sec:epent}

As  recalled above in Sec.~2.1, globally pure quantum states are entangled when their projections onto local states (performed by partial traces) are mixed states. We will now employ the same argument, properly adapted, for classical dynamical systems with coarse-graining.

\subsection{Definition}
\label{sec:epentdef}

In order to so, we use the notion of an epistemically pure state of Section \ref{sec:psp}: A statistical state $\rho$ is called \emph{epistemically pure with respect to an observable $f$} if $\rho$ is an eigenstate of $f$ and if $\rho$ is a  finest-refined state epistemically accessible by $f$.

We now define a modified kind of entanglement for classical systems in the following way: A state $\rho$ is \emph{epistemically entangled with respect to an observable $f$} if $\rho$ is epistemically pure with respect to a global observable $f$ and if $\rho$ is not epistemically pure with respect to some local observable $g$. That means that $\rho$ would appear as a mixture after projection into a local subsystem.

\subsection{Example: Coupled Harmonic Oscillators}
\label{sec:epentcho}

Consider two coupled harmonic oscillators obeying
\begin{eqnarray}
 \label{eq:copossi1}
  \ddot{q}_1 + \omega^2 q_1 &=& - \frac{\eta^2}{2}(q_1 - q_2) \:, \\
  \label{eq:copossi2}
  \ddot{q}_2 + \omega^2 q_2 &=& \frac{\eta^2}{2}(q_1 - q_2) \:,
\end{eqnarray}
with positions $q_1, q_2$, frequency $\omega$ and coupling constant $\eta$. The system of equations (\ref{eq:copossi1}, \ref{eq:copossi2}) can be decoupled using normal coordinates $u_1 =q_1 - q_2$ and
$u_2 = q_1 + q_2$, leading to
\begin{eqnarray}
 \label{eq:copossinc1}
  \ddot{u}_1 + \tilde{\omega}^2 u_1 &=& 0 \:, \\
  \label{eq:copossinc2}
  \ddot{u}_2 + \omega^2 u_2 &=& 0 \:,
\end{eqnarray}
with normal frequencies $\omega$ and $\tilde{\omega} = \sqrt{\omega^2 + \eta^2}$.

Each equation is independently solved by (\ref{eq:dglsolv1}, \ref{eq:dglsolv2}), such that in the four-dimensional phase space, spanned by $x = (u_1, u_2, v_1, v_2)^T$, the flow is given by \begin{equation}\label{eq:normflow}
    \tilde{\Phi}^t =
    \begin{pmatrix}
      \cos \tilde{\omega} t & 0 & \frac{1}{\tilde{\omega}} \sin \tilde{\omega} t & 0 \\
      0 & \cos \omega t & 0 & \frac{1}{\omega} \sin \omega t \\
      -\tilde{\omega} \sin \tilde{\omega} t & 0 & \cos \tilde{\omega} t & 0 \\
      0 & -\omega \sin \omega t & 0 & \cos \omega t \\
    \end{pmatrix} \:,
\end{equation}
and $x(t) = \tilde{\Phi}^t \cdot x(0)$, where $v_1 = \dot{u}_1$, $v_2 = \dot{u}_2$ are the canonically conjugate momenta.

Now we assume that the normal frequencies are commensurable. This can easily be achieved by letting $\eta^2 = 3 \omega^2$, leading to $\tilde{\omega} = 2 \omega$, and thus $\tilde{\omega} : \omega = 2 : 1$. Under this crucial simplification also the periods of the normal oscillatory modes are commensurable: $T_1 = T_2 / 2$.  If we now choose a sampling period $\tau = T_1$, the discretized flow becomes \begin{equation}\label{eq:normflowsmp}
    \tilde{\Phi}^\tau =
    \begin{pmatrix}
      1 & 0 & 0 & 0 \\
      0 & -1 & 0 & 0 \\
      0 & 0 & 1 & 0 \\
      0 & 0 & 0 & -1 \\
    \end{pmatrix} \:.
\end{equation}

As for a single harmonic oscillator (Section 3.2), we have $(\tilde{\Phi}^\tau)^2 = \eins$ and, therefore, $(\tilde{\Phi}^\tau)^{-1} =\tilde{\Phi}^{-\tau} = \tilde{\Phi}^\tau$, and thus
\begin{equation}\label{eq:invflow}
 \tilde{\Phi}^{-\tau} \cdot
 \begin{pmatrix}
   u_1 \\
   u_2 \\
   v_1 \\
   v_2 \\
 \end{pmatrix} =
 \begin{pmatrix}
   u_1 \\
   -u_2 \\
   v_1 \\
   -v_2 \\
 \end{pmatrix} \:.
\end{equation}

Next, we introduce four  coarse-grained \emph{global observables} $U_i(x) = \Theta(u_i)$, $V_i(x) = \Theta(v_i)$ ($i = 1, 2$), each bi-partitioning one normal axis. Consider, e.g., the cells $D_0 = \{ x \in X | U_2(x) = 0 \}$ and $D_1 = \{ x \in X | U_2(x) = 1 \}$ which belong to a partition $\mathcal{U}_2$. Then we have $\tilde{\Phi}^{-\tau}(D_0) = D_1$ and, vice versa, $\tilde{\Phi}^{-\tau}(D_1) = D_0$, such that $\mathbf{R} \mathcal{U}_2 = \mathcal{U}_2$. Expressed in terms of the original phase space coordinates, the cell $D_1$ (which is given by $u_2 \ge 0$) is retained as $q_1 \ge -q_2$. The eigenstate associated  with
\begin{equation}\label{eq:globeigenstate}
    p(x) = 2 \chi_{D_1}(x)
\end{equation}
is a finest epistemically accessible state with respect to $U_2$ and, therefore, epistemically pure with respect to $U_2$.

In order to obtain \emph{local observables}, we transform the normal flow $\tilde{\Phi}^t$ back into local coordinates $x = (q_1, q_2, p_1, p_2)^T$. This is achieved by a mixing matrix
\begin{equation}\label{eq:mixmat}
    M = \frac{1}{2}
    \begin{pmatrix}
      1 & 1 & 0 & 0 \\
      -1 & 1 & 0 & 0 \\
      0 & 0 & 1 & 1 \\
      0 & 0 & -1 & 1 \\
    \end{pmatrix} \:,
\end{equation}
such that $\Phi^t = M \cdot \tilde{\Phi}^t \cdot M^{-1}$. For  the sampling period $\tau = T_1$ we obtain
\begin{equation}\label{eq:locflow}
    \Phi^\tau =
    \begin{pmatrix}
      0 & -1 & 0 & 0 \\
      -1 & 0 & 0 & 0 \\
      0 & 0 & 0 & -1 \\
      0 & 0 & -1 & 0 \\
    \end{pmatrix} = \Phi^{-\tau} \:,
\end{equation}
such that
\begin{equation}\label{eq:invflowloc}
 \Phi^{-\tau} \cdot
 \begin{pmatrix}
   q_1 \\
   q_2 \\
   p_1 \\
   p_2 \\
 \end{pmatrix} =
 \begin{pmatrix}
   -q_2 \\
   -q_1 \\
   -p_2 \\
   -p_1 \\
 \end{pmatrix} \:.
\end{equation}

Now we can evaluate the effect of this flow upon  the four coarse-grained local observables $Q_i(x) = \Theta(q_i)$, $P_i(x) = \Theta(p_i)$ ($i = 1, 2$). Let $\mathcal{Q}_1 = \{ G_0, G_1 \}$ and $\mathcal{Q}_2 = \{ H_0, H_1 \}$ be the partitions induced by $Q_1$ and $Q_2$, respectively. Then, $\Phi^{-\tau}(G_0) = H_1$, $\Phi^{-\tau}(G_1) = H_0$, $\Phi^{-\tau}(H_0) = G_1$, $\Phi^{-\tau}(H_1) = G_0$, such that $\mathbf{R} \mathcal{Q}_1 = \mathbf{R} \mathcal{Q}_2 = \mathcal{Q}_1 \vee \mathcal{Q}_2$.

Consider again the eigenstate $\rho$ associated  with $p$ from \Eq{eq:globeigenstate}. This p.d.f.~is not measurable by the partition algebra $\mathcal{B}_{Q_1}$ and hence not epistemically accessible with respect to the local observable $Q_1$. Therefore, while $\rho$ is epistemically pure with respect to the global observable $U_2$, it is not epistemically pure with respect to the local observable $Q_1$.  Thus, $\rho$ is an epistemically entangled state.

\section{Conclusions}
\label{sec:disc}

Measurements of classical systems typically refer to representations of the state of the system in a coarse-grained phase space. In other words, epistemically equivalent states partition the phase space and introduce an ``epistemic quantization''. The distinguished class of generating partitions allows us to approximate individual states by continuous measurements, leading to a dynamical refinement of the partition. Only in this special case will individual, pointwise represented states be epistemically accessible.

However, the general situation of non-generating, ``improper'' partitions is also interesting. In previous work \citep{GrabenAtmanspacher06a} we showed that classical observables can be \emph{complementary} with respect to particular partitions if not all states in the full state space are epistemically accessible. In the same spirit, this paper argues that the notion of entanglement can be employed for classical dynamical systems.

The strategy to show this rests on properly defined ``epistemically pure'' dispersive classical states, in contrast to ``ontically pure'' dispersive quantum states. We demonstrated that, under particular conditions, classical states that are epistemically pure with respect to a global observable may be mixed with respect to a suitably defined local observable.

The structure of this argument resembles the definition of quantum entanglement, where the pure state of a system as a whole is not separable into pure states of decomposed subsystems. The decomposition leads to mixed states, indicating correlations between the subsystems. We demonstrated for the example of coupled harmonic oscillators how a related inseparability occurs for classical dynamical systems with non-generating partitions.

This result substantiates a recent conjecture of epistemic entanglement \citep{AtmFilkGrab11}. It should be of interest to all kinds of applications which can be formalized in terms of phase space represenations. This applies in particular to situations in cognitive science for which phase space representations have generated increasing attention \citep{Gelder98}. Recent work by \citet{Bruza09} on non-separable concept combinations in the human mental lexicon are an especially interesting example in this respect.

\section*{Acknowledgements}

This research was partly supported by the Franklin Fetzer Fund and by a DFG Heisenberg grant awarded to PbG (GR 3711/1-1).




\end{document}